\documentclass{article}
\usepackage{piner}
\usepackage[figuresright]{rotating}
\begin{document}
\submitted{Submitted to ApJ, Aug. 5, 2003}

\title{The Parsec-Scale Structure and Jet Motions of the TeV Blazars
1ES~1959+650, PKS~2155$-$304, and 1ES~2344+514}

\author{B. Glenn Piner\altaffilmark{1,2} and Philip G. Edwards\altaffilmark{3}}

\altaffiltext{1}{Department of Physics and Astronomy, Whittier College,
13406 E. Philadelphia Street, Whittier, CA 90608; gpiner@whittier.edu}

\altaffiltext{2}{NASA Summer Faculty Fellowship, Jet Propulsion Laboratory,
California Institute of Technology, 4800 Oak Grove Drive, Pasadena, CA
91109}

\altaffiltext{3}{Institute of Space and Astronautical Science, Yoshinodai, Sagamihara, Kanagawa 229-8510, Japan;
pge@vsop.isas.ac.jp}

\begin{abstract}
As part of our study of the VLBI properties of TeV blazar jets,
we present here a series of high-resolution 15 GHz Very Long Baseline Array (VLBA) images of the parsec-scale jets
of the TeV blazars 1ES~1959+650, PKS~2155$-$304, and 1ES~2344+514, with linear resolutions of $\sim0.5$ pc.
Each of these sources was observed with the VLBA at three or four epochs during 1999 and 2000.
There is a notable lack of any strong moving components on the VLBI images
(in contrast to the rapid superluminal motions seen in EGRET blazars),
and the structure of the VLBI jet can be modeled either as a series of stationary Gaussian components,
or as a smooth power law for two of the sources (PKS~2155$-$304 and 1ES~2344+514).
The low apparent speeds, together with beaming
indicators such as the brightness temperature of the VLBI core, imply only modest Doppler boosting
of the VLBI radio emission, and only modest bulk Lorentz factors ($\delta$ and $\Gamma\approx$ a few); in contrast to the more
extreme values of these parameters invoked to explain the high-energy emission.
The fact that no moving shocks or plasmoids are seen on the parsec-scale 
suggests that the shocks or plasmoids that are assumed to be responsible for the high-energy flares
must dissipate before they separate from the core on the VLBI images.
This requires the loss of a substantial amount of bulk kinetic energy on parsec scales, and
implies a higher efficiency than is typically assumed for internal shock scenarios.
\end{abstract}


\section{Introduction}
\label{intro}
The blazar phenomenon is well established to be the result of relativistic electrons
(and possibly positrons) radiating in a jet which is undergoing bulk relativistic motion
at a small angle to the observer's line-of-sight.
The characteristic blazar spectral energy distribution is two-peaked, with the low-frequency peak due to synchrotron
radiation, and the high-frequency peak due to inverse-Compton scattering, either
of the jets own synchrotron photons (synchrotron self-Compton, or SSC emission), or of external photons
(the so-called external-radiation Compton, or ERC emission).
In VLBI images, most blazars display rapid superluminal apparent motions
(Jorstad et al. 2001a).
A particular sub-class of blazar whose synchrotron spectra in a $\nu F_{\nu}$ plot peak at X-ray frequencies
(the high-frequency peaked BL Lac objects, or HBLs)
has inverse-Compton spectra that peak at TeV $\gamma$-ray frequencies, and a few of these objects have been detected
by ground-based TeV $\gamma$-ray telescopes (Horan et al. 2002).
This set of objects is referred to as the TeV blazars.
The TeV blazars are restricted to relatively nearby HBLs, because of the absorption of TeV $\gamma$-rays
by pair-production on the extragalactic background light.

The TeV blazars display a number of interesting characteristics:
They show dramatic variability in their high-energy emission (e.g., Gaidos et al. 1996), 
and the size scales implied by such rapid variability require large
relativistic Doppler factors 
in the $\gamma$-ray producing region, to avoid $\gamma$-ray absorption
(the Doppler factor $\delta=1/\Gamma(1-\beta\cos\theta)$, where
$\Gamma$ is the bulk Lorentz factor, $\beta=v/c$, and $\theta$ is the angle to the line-of-sight).
Some specific emission models require extreme Doppler factors and bulk Lorentz factors
($\sim$ 40-50); these values can be somewhat reduced in some models but remain high
($\sim$ 15) (Georganopoulos \& Kazanas 2003).  
The TeV blazar phenomenon thus requires both high particle and bulk Lorentz factors.
The high particle Lorentz factors that are required to produce the high-frequency emission
are commonly assumed to be produced at shocks in the jet, where some of the bulk kinetic
energy of the flow is converted to internal kinetic energy.  These shocks may be produced either by internal interactions
in the jet (e.g., Spada et al. 2001), or by interaction of the jet with the external environment (e.g., Dermer \& Chiang 1998).

This paper is part of our ongoing program to study the parsec-scale jets
of a TeV $\gamma$-ray-selected sample of blazars using the National Radio Astronomy Observatory's
Very Long Baseline Array (VLBA)\footnote{The National Radio Astronomy Observatory is a facility of the National
Science Foundation operated under cooperative agreement by Associated Universities, Inc.}.
VLBI observations of these TeV sources provide the
highest-resolution images of these jets currently achievable.  
These images provide independent constraints on parameters that are crucial 
for modeling the jet.  Observations of changes
in jet structure on parsec scales (superluminal motion) can constrain both the Lorentz factor of the jet and the angle of the jet to
the line-of-sight (the apparent jet speed $\beta_{app}=\beta\sin\theta/(1-\beta\cos\theta)$), 
and also the jet opening angle.  
In addition, there is the possibility of directly imaging the shock,
or ``component'', responsible for the $\gamma$-ray flare emission, as has apparently been done for some
EGRET blazars (Jorstad et al. 2001b).
The VLBI observations probe the inner-jet properties through direct imaging, and these properties
can then be compared with those deduced from light curves and multiwavelength spectra.

There are currently five extragalactic sources whose detection in TeV $\gamma$-rays
has been confirmed by multiple detections, all of these are HBLs.
They are: Markarian~421, Markarian~501,
1ES~1959+650, and H~1426+428 (Horan et al. 2003 and references therein),
and PKS~2155$-$304 which has recently been confirmed (Djannati-Ata\"{\i} et al. 2003).  
In addition, there have been unconfirmed detections
of five other sources: the HBL 1ES~2344+514,
the low-frequency peaked BL Lac objects BL Lac and 3C~66A (Horan et al. 2002 and references therein),
the radio galaxy M87 (Aharonian et al. 2003a), and the starburst galaxy NGC~253 (Itoh et al. 2002).
The confirmed detections of five HBLs make the TeV emission of this class of object well established,
and we therefore restrict our sample to the six TeV-detected HBLs (the five confirmed sources mentioned above
plus 1ES~2344+514), until such time as the TeV emission from one of the other classes
of candidate objects is confirmed.

Our earlier observations of Mrk~421 were presented by Piner et al. (1999); more recent
VLBA observations following the prolonged flaring state of Mrk~421 in early 2001
will be presented by Piner \& Edwards, in preparation.  Observations of Mrk~501 were reported
by Edwards \& Piner (2002), see also Giroletti et al. (2003).  
This paper describes a series of VLBA observations of 1ES~1959+650,
PKS~2155$-$304, and 1ES~2344+514 obtained during 1999 and 2000.  Initial results on these three sources
were briefly presented by Piner et al. (2002), here we provide more detailed information and a fuller description
and consideration of the entire dataset.  Because H~1426+428 was not detected as a TeV source until relatively
recently, it was not added to our observing campaign until 2001.  Observations of H~1426+428, plus
additional monitoring of 1ES~1959+650 and PKS~2155$-$304, are underway, and will be presented in a future paper.
The VLBA flux densities of the three sources studied in this paper are substantially less ($\sim$ 100 mJy)
than those of the two more well-studied TeV blazars Mrk~421 and Mrk~501 ($\sim$ 500 mJy).

In this paper we use the cosmological parameters measured by the Wilkinson Microwave Anisotropy Probe (WMAP) of
$H_{0}=71$ km s$^{-1}$ Mpc$^{-1}$, $\Omega_{m}=0.27$, and $\Omega_{\Lambda}=0.73$
(Bennett et al. 2003).  We use the equation given in footnote (14) of Perlmutter et al. (1997)
to compute the luminosity distance for non-zero $\Omega_{\Lambda}$. 
When results from other papers are quoted,
these results have been converted to the set of cosmological parameters given above.
In $\S$~\ref{sources} we give a more detailed introduction to the three sources in this paper,
in $\S$~\ref{obs} and $\S$~\ref{results} we discuss the VLBI observations and the results obtained
from imaging and model fitting, and in $\S$~\ref{discussion} we discuss the astrophysical implications of these results.

\section{The Individual Sources}
\label{sources}
\subsection{1ES~1959+650}
1ES~1959+650 ($z=0.047$) became the third source to have its TeV emission confirmed, after Mrk~421 and Mrk~501.
It was initially detected at a low significance in 1998 by the Utah Seven Telescope Array (Nishiyama et al. 2000),
and it has subsequently been detected by the Whipple (Holder et al. 2003a), HEGRA (Aharonian et al. 2003b),
and CAT (Djannati-Ata\"{\i} et al. 2002) experiments.
The highest levels of TeV $\gamma$-ray activity were observed between 2002 May and 2002 July, when the source
showed two strong flares, the strongest
peaking at a flux of 5 crab with a doubling time of 7 hours (Holder et al. 2003a).
TeV $\gamma$-ray observations at other times (2000-2001 and after 2002 September) have shown
the source to be in a relatively quiescent state, with an average flux of only 0.05 crab in 2000-2001
(Aharonian et al. 2003b; Holder et al. 2003b).  1ES~1959+650 is also quite variable at X-ray
(Giebels et al. 2002; Holder et al. 2003b) and optical (Villata et al. 2000) wavelengths.
A multiwavelength spectrum is shown by Beckmann et al. (2002).  A 5 GHz VLBA image is given by Bondi et al. (2001),
and 5 GHz VLBA and 1.4 GHz VLA images
are presented by Rector, Gabuzda, \& Stocke (2003).

\subsection{PKS~2155$-$304}
As an archetypal X-ray-selected BL Lac object, PKS~2155$-$304 ($z=0.117$) has been extensively
observed by X-ray and UV satellites, and ground-based optical
monitoring, and it exhibits complex phenomenology in its rapid,
strong, broadband variability.  Recent results include descriptions of {\em BeppoSAX} data by Zhang et al. (2002),
{\em XMM-Newton} data by Edelson et al. (2001), optical polarimetric monitoring by Tommasi et al. (2001),
and a long-look {\em ASCA} observation by Tanihata et al. (2001).
This source was detected in TeV $\gamma$-rays in late 1996 and late 1997
by the University of Durham Mark 6 telescope, with a combined significance of 6.8$\sigma$
(Chadwick et al. 1999).  The peak TeV flux was measured in 1997 November, when PKS~2155$-$304 was in an
X-ray high-state (Chadwick et al. 1999; Chiappetti et al. 1999).  
A confirming detection of this source in TeV $\gamma$-rays has been made recently 
by the H.E.S.S. collaboration (Djannati-Ata\"{\i} et al. 2003).  Specific emission models 
used to explain the spectra and variability of this source have used Doppler factors
ranging from $\delta\sim 20-30$ (Chiappetti et al. 1999; Kataoka et al. 2000); however, the
observations on which these results are based have been disputed by Edelson et al. (2001).
Radio observations of PKS~2155$-$304 are sparse compared to the optical and high-energy observations,
but a sequence of VLA images is presented by Laurent-Muehleisen et al. (1993).

\subsection{1ES~2344+514}
1ES~2344+514 ($z=0.044$) was detected in TeV $\gamma$-rays by the Whipple telescope (Catanese et al. 1998)
during an active period from 1995 October to 1996 January, when it displayed one flare with a
flux of 0.6 crab (a 6$\sigma$ detection), and an average flux level of 0.1 crab (excluding the flare, a
4$\sigma$ detection).
Although subsequent TeV observations of this source have yielded only upper limits (e.g., Aharonian et al. 2000),
it is the most secure of the unconfirmed detections, in part because
1ES~2344+514 is expected to be one of the brightest extragalactic TeV sources (Ghisellini 2003).
Detection of exceptional X-ray spectral variability is reported by Giommi, Padovani, \& Perlman (2000)
from {\em BeppoSAX} observations between 1996 and 1998.  They observe shifts by a factor of 30 or more
in the peak frequency of the synchrotron emission, which ranged to at or above 10 keV.  Rapid variability
on time-scales of 5000s was also detected when the source was brightest.
Detection of more moderate variability at optical wavelengths is discussed by Xie et al. (2002).
VLBI images at 1.6 and 5 GHz are presented by Bondi et al. (2001), and 
5 GHz VLBA and 1.4 GHz VLA images are shown by Rector et al. (2003).

\section{Observations}
\label{obs}
\subsection{Details of Observations}
We observed 1ES~2344+514 with the VLBA at 15 GHz at four epochs between 1999 October and 2000 March
with 8 hours of observation time per epoch (a total of 32 hours),
under observation code BP057.  We observed 1ES~1959+650 and PKS~2155$-$304 at 15 GHz at three epochs
each between 2000 March and 2000 July with 6 hours of observation time per source per epoch (a total of 36 hours),
under observation code BP062.  All of the observations used standard VLBA continuum setups
(8 intermediate frequencies, 64 MHz total bandwidth, 1-bit sampling for BP057; 
4 intermediate frequencies, 32 MHz total bandwidth, 2-bit sampling
for BP062), and all recorded left circular polarization.
An observation log is given in Table~\ref{imtab}.
An observing frequency of 15 GHz
was chosen because it provided the best combination of the high-resolution needed to monitor possible moving
jet components with the sensitivity required to image jets in these relatively faint $\sim$ 100 mJy
sources with adequate dynamic range.
Calibration and fringe-fitting were done with
the AIPS software package.  Images from these datasets were produced using standard CLEAN and
self-calibration procedures from the Difmap software package (Shepherd, Pearson, \& Taylor 1994).
The size of the VLBA beam at 15 GHz is approximately 0.5 milliarcseconds (mas); this corresponds to linear resolutions
of about 0.5 pc for 1ES~1959+650 and 1ES~2344+514, and about 1 pc for PKS~2155$-$304.

\begin{table*}
\caption{Observation Log and Parameters of the Images}
\label{imtab}
{\tiny \begin{tabular}{l l c l r c c r c c} \tableline \tableline \\
& & & & \multicolumn{3}{c}{Natural Weighting} & \multicolumn{3}{c}{Uniform Weighting} \\ \tableline \\
& & Time on & & & Peak Flux & Lowest & & Peak Flux & Lowest \\
& & Source  & \multicolumn{1}{c}{VLBA} & & Density & Contour$^{b}$ & & Density & Contour$^{b}$ \\
Source & Epoch & (hours) & \multicolumn{1}{c}{Antennas} & \multicolumn{1}{c}{Beam$^{a}$} &
(mJy beam$^{-1}$) & (mJy beam$^{-1}$) & \multicolumn{1}{c}{Beam$^{a}$}
& (mJy beam$^{-1}$) & (mJy beam$^{-1}$) \\ \tableline \\
1ES~1959+650 & 2000 Mar 6  & 6 & No KP,OV & 0.95,0.47,7.6     & 96  & 0.25 & 0.74,0.33,9.5    & 85  & 0.45 \\ [5pt]
             & 2000 Jun 9  & 6 & All      & 0.95,0.51,10.7    & 96  & 0.23 & 0.70,0.36,15.4   & 82  & 0.44 \\ [5pt]
             & 2000 Jul 8  & 6 & No PT    & 0.91,0.49,9.4     & 98  & 0.28 & 0.68,0.35,15.6   & 85  & 0.50 \\ [5pt]
PKS~2155$-$304 & 2000 Mar 3  & 6 & All      & 1.38,0.49,$-$1.1  & 210 & 0.34 & 1.01,0.35,$-$2.0 & 194 & 0.62 \\ [5pt]
             & 2000 Jun 2  & 6 & No NL    & 1.49,0.49,$-$3.4  & 137 & 0.33 & 0.99,0.34,$-$0.5 & 124 & 0.95 \\ [5pt]
             & 2000 Jun 29 & 6 & No FD    & 1.41,0.48,$-$5.5  & 155 & 0.38 & 1.00,0.35,$-$1.8 & 143 & 0.82 \\ [5pt]
1ES~2344+514 & 1999 Oct 1  & 8 & All      & 0.91,0.56,$-$11.6 & 96  & 0.17 & 0.62,0.37,$-$4.8 & 91  & 0.43 \\ [5pt]
             & 1999 Nov 9  & 8 & All      & 0.91,0.55,$-$10.7 & 93  & 0.18 & 0.63,0.37,$-$4.7 & 89  & 0.40 \\ [5pt]
             & 2000 Jan 7  & 8 & All      & 0.92,0.56,$-$11.3 & 78  & 0.17 & 0.63,0.37,$-$4.0 & 73  & 0.39 \\ [5pt]
             & 2000 Mar 23 & 8 & No PT    & 0.84,0.51,$-$7.9  & 90  & 0.20 & 0.62,0.36,$-$4.0 & 85  & 0.37 \\ \tableline
\end{tabular}}
\\
{\scriptsize $a$: Numbers given for the beam are the FWHMs of the major
and minor axes in mas, and the position angle of the major axis in degrees.
\\ Position angle is measured from north through east.}\\
{\scriptsize $b$: The lowest contour is set to be three times the rms noise
in the image. Successive contours are each a factor of 2 higher.}
\end{table*}

\subsection{Images}
The VLBA images of 1ES~1959+650, PKS~2155$-$304, and 1ES~2344+514 are shown in Figures 1 to 3, 
respectively.  In each of these figures, the top row contains the higher-sensitivity but
lower-resolution naturally-weighted images (uvweight=0,$-$2 in Difmap), while the bottom row
contains the lower-sensitivity but higher-resolution uniformly-weighted images 
(uvweight=2,0 in Difmap).  Parameters of these images are given in Table~\ref{imtab}.
The rms noise in the images is approximately equal to the thermal noise limit for 
1ES~2344+514 and 1ES~1959+650, for PKS~2155$-$304 it is about 30\% higher.
The dynamic range (peak/rms) achieved exceeds 1000:1 at all epochs for the naturally-weighted images. 

\begin{sidewaysfigure*}
\plotfiddle{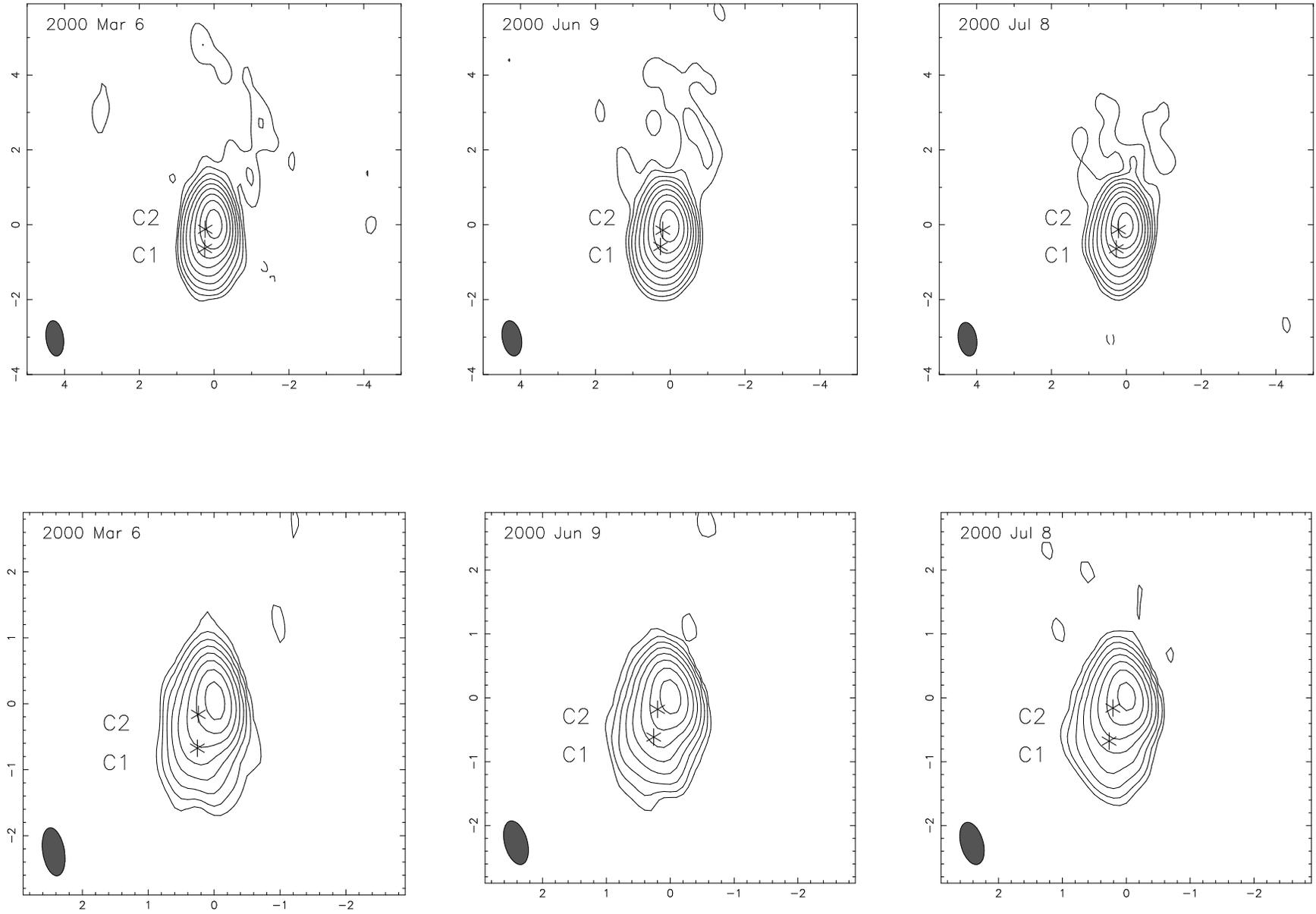}{6.5in}{-90}{90}{90}{-363}{500}
\caption{VLBA images of 1ES~1959+650 at 15 GHz.
The epoch of observation is shown in the upper-left corner of each image.
The top row shows the images obtained with natural weighting (uvweight=0,-2 in DIFMAP),
the bottom row shows the images obtained with uniform weighting (uvweight=2,0 in DIFMAP).
The axes are labeled in milliarcseconds.
Numerical parameters of the images are given in Table~\ref{imtab}.
The centers of the circular Gaussians (excluding the Gaussian representing the core) that were fit to
the visibilities are marked with asterisks, and the identifications of the Gaussians are indicated.
Numerical parameters of these Gaussians are given
in Table~\ref{mfittab}.}
\end{sidewaysfigure*}

\begin{sidewaysfigure*}
\plotfiddle{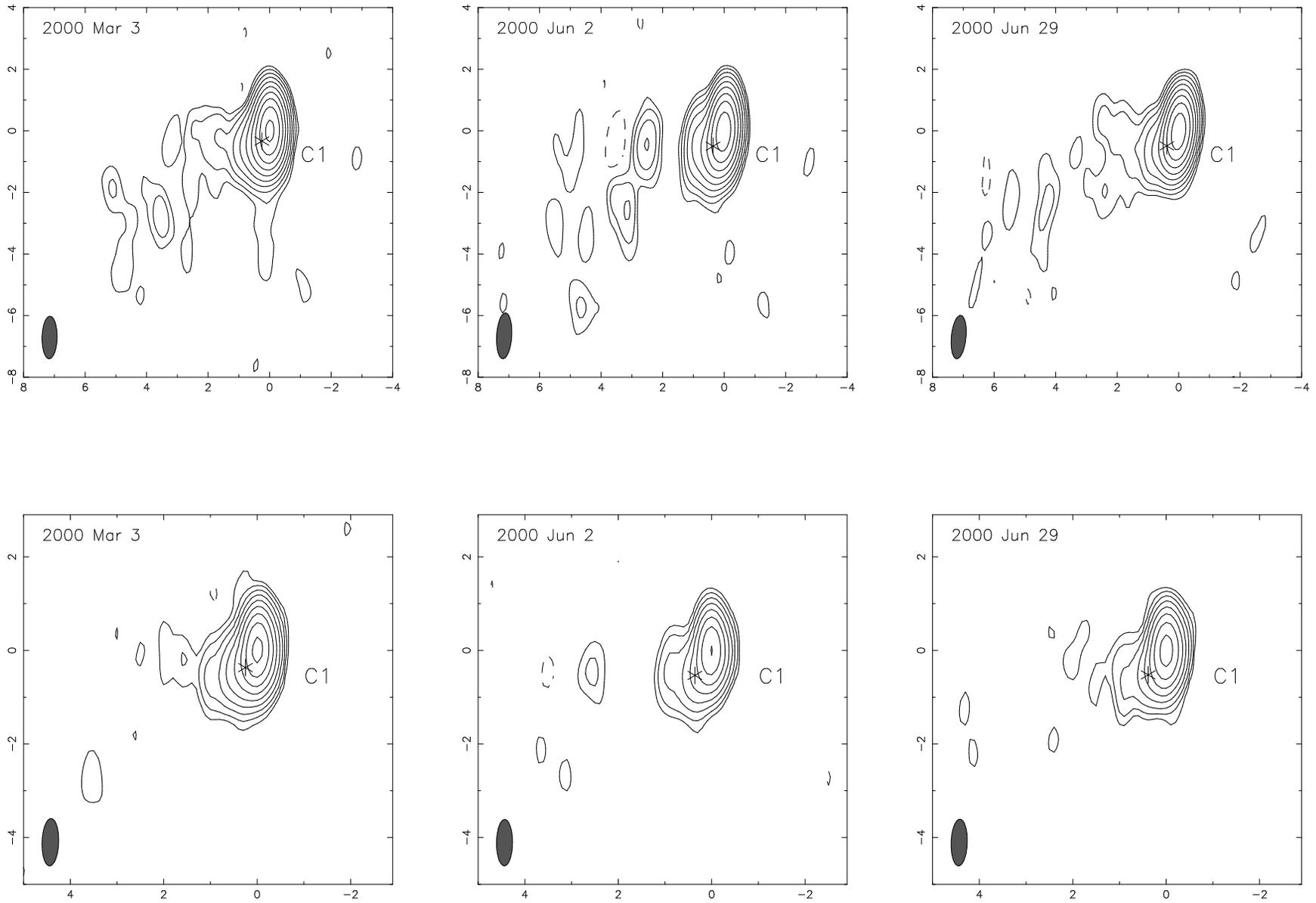}{6.5in}{-90}{90}{90}{-363}{500}
\caption{VLBA images of PKS~2155$-$304 at 15 GHz.
The epoch of observation is shown in the upper-left corner of each image.
The top row shows the images obtained with natural weighting (uvweight=0,-2 in DIFMAP),
the bottom row shows the images obtained with uniform weighting (uvweight=2,0 in DIFMAP).
The axes are labeled in milliarcseconds.
Numerical parameters of the images are given in Table~\ref{imtab}.
The centers of the circular Gaussians (excluding the Gaussian representing the core) that were fit to
the visibilities are marked with asterisks, and the identifications of the Gaussians are indicated.
Numerical parameters of these Gaussians are given
in Table~\ref{mfittab}.}
\end{sidewaysfigure*}

\begin{sidewaysfigure*}
\plotfiddle{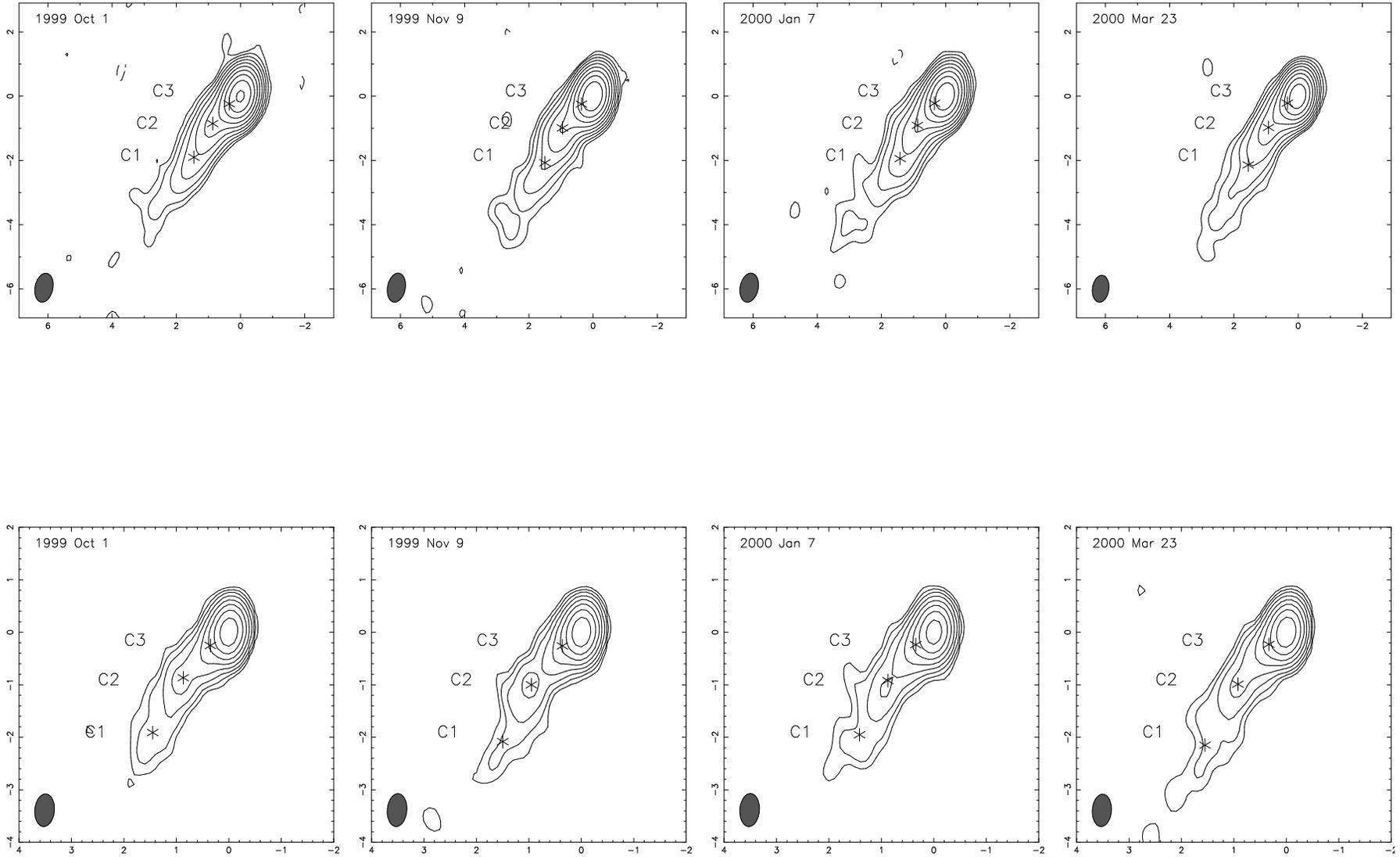}{6.5in}{-90}{90}{90}{-363}{500}
\figcaption{VLBA images of 1ES~2344+514 at 15 GHz.
The epoch of observation is shown in the upper-left corner of each image.
The top row shows the images obtained with natural weighting (uvweight=0,-2 in DIFMAP),
the bottom row shows the images obtained with uniform weighting (uvweight=2,0 in DIFMAP).
The axes are labeled in milliarcseconds.
Numerical parameters of the images are given in Table~\ref{imtab}.
The centers of the circular Gaussians (excluding the Gaussian representing the core) that were fit to
the visibilities are marked with asterisks, and the identifications of the Gaussians are indicated.
Numerical parameters of these Gaussians are given
in Table~\ref{mfittab}.}
\end{sidewaysfigure*}

1ES~1959+650 has an intriguing morphology.
The uniformly-weighted images in Figure~1 clearly show a short jet extending 1 mas to the
southeast of the core (presumed to be the northernmost of the three components we see,
because it is both the brightest and the most compact), along a position angle of $\approx 160\arcdeg$.
The naturally-weighted images also show this southeastern jet, but in addition show 
broad, diffuse emission to the north of the core.
The lower-resolution 5 GHz VLBA image of 1ES~1959+650 by Rector et al. (2003) 
(see also the similar image by Bondi et al. 2001) shows
a diffuse jet with a broad ($\sim 55\arcdeg$) opening angle extending 20 mas north
of the core along a position angle of $\approx -5\arcdeg$.  There is no indication of 
a southern jet in those image, but the southern jet seen in this paper would not be resolved 
by those observations.  
The VLA image of Rector et al. (2003) shows faint
extended flux to the north (P.A. $\approx -5\arcdeg$) and south (P.A. $\approx 175\arcdeg$) of the core. 
Because no southern jet is visible in the lower-resolution 5 GHz image, we do
not think the VLBA images are showing a jet and counterjet, but instead speculate that the jet from
this source may be very closely aligned with the line-of-sight, such that a slight bend
carries the jet across the line-of-sight and causes it to appear first to the south of
the core in the 15 GHz VLBA images, and then to the north of the core in the 5 GHz VLBA image. 

PKS~2155$-$304 has a jet that starts to the southeast of the core at a position angle of $\approx 150\arcdeg$,
before bending toward the east at about 1 mas from the core.  At that point the jet becomes broader
and more diffuse, before finally bending again to the southeast at about 3 mas from the core.
As far as we know, these are the first published VLBI images of this source.
The highest-resolution VLA image of PKS~2155$-$304 presented by Laurent-Muehleisen et al. (1993) 
shows a knot nearly 180$\arcdeg$ misaligned from the VLBA jet in Figure~2,
the lower-resolution VLA images by these same authors also show an extended halo of emission around the core.
Both 1ES~1959+650 and PKS~2155$-$304 show extreme jet misalignments, either in their
parsec-scale structure or between the parsec and kiloparsec scales.

1ES~2344+514 shows a typical core-jet morphology, with a nearly straight jet along a position angle
of $\approx 145\arcdeg$, detectable out to 4 mas from the core in our images.  The 5 GHz VLBA
image of Rector et al. (2003) (see also the similar image by Bondi et al. 2001)
shows this jet extending to 50 mas from the core, but becoming
more diffuse and broadening into a cone with a $\sim 35\arcdeg$ opening angle.
The VLA image of Rector et al. (2003) detects emission extending to the east
(P.A. $\approx 105\arcdeg$) in a 50$\arcdeg$ cone.  The misalignment angle for this source
is about 40$\arcdeg$.

\subsection{Model Fits}
\label{modelfits}
In order to quantify any possible motions in these jets, circular Gaussian model components
were fit to the visibility data using the ``modelfit'' task in Difmap.
The parameters of these model fits are given in Table~\ref{mfittab}.
The reduced $\chi^{2}$ for all model fits was under 1.0.
1ES~1959+650 was well fit by two circular Gaussians (C1 and C2) in addition to the core component.
These components are located in the southeastern jet visible in the uniformly-weighted images
of Figure~1.  C1 is a $\sim 20$ mJy component located about 0.8 mas from the core,
C2 is a $\sim 30$ mJy component located about 0.4 mas from the core.
The visibilities for PKS~2155$-$304 are well fit by only a single circular Gaussian in addition
to the core, with a flux density $\sim 40$ mJy and a core separation of $\approx 0.6$ mas.
The jet in 1ES~2344+514 is fit by three Gaussians (C1, C2, and C3).
The flux densities of C1, C2, and C3 are $\sim$ 5, 10, and 15 mJy, and the core separations are
$\approx$ 2.5, 1.3, and 0.5 mas, respectively.
The centers of the model-fit Gaussian components are marked by asterisks on Figures 1 to 3.

\begin{table*}[!t]
\begin{center}
{\small
\caption{Circular Gaussian Models}
\label{mfittab}
\begin{tabular}{l l c c c c c} \tableline \tableline \\ [-5pt]
& & & $S$\tablenotemark{a} & $r$\tablenotemark{b} &
PA\tablenotemark{b} & $a$\tablenotemark{c} \\ [5pt]
Source & Epoch & Component & (mJy) & (mas) &
(deg) & (mas) \\ \tableline \\ [-5pt]
1ES~1959+650 & 2000 Mar 6  & Core & 91  & ...  & ...   & 0.14 \\ [5pt]
             &             & C2   & 25  & 0.37 & 134.0 & 0.29 \\ [5pt]
             &             & C1   & 18  & 0.82 & 159.9 & 0.56 \\ [5pt]
             & 2000 Jun 9  & Core & 83  & ...  & ...   & 0.15 \\ [5pt]
             &             & C2   & 31  & 0.35 & 140.5 & 0.27 \\ [5pt]
             &             & C1   & 22  & 0.76 & 158.1 & 0.52 \\ [5pt]
             & 2000 Jul 8  & Core & 85  & ...  & ...   & 0.14 \\ [5pt]
             &             & C2   & 36  & 0.35 & 137.1 & 0.27 \\ [5pt]
             &             & C1   & 19  & 0.83 & 159.0 & 0.59 \\ [5pt]
PKS~2155$-$304 & 2000 Mar 3  & Core & 205 & ...  & ...   & 0.18 \\ [5pt]
             &             & C1   & 52  & 0.53 & 150.2 & 0.75 \\ [5pt]
             & 2000 Jun 2  & Core & 138 & ...  & ...   & 0.18 \\ [5pt]
             &             & C1   & 33  & 0.70 & 149.3 & 0.53 \\ [5pt]
             & 2000 Jun 29 & Core & 165 & ...  & ...   & 0.21 \\ [5pt]
             &             & C1   & 26  & 0.71 & 146.1 & 0.68 \\ [5pt]
1ES~2344+514 & 1999 Oct 1  & Core & 93  & ...  & ...   & 0.08 \\ [5pt]
             &             & C3   & 13  & 0.46 & 129.0 & 0.23 \\ [5pt]
             &             & C2   & 10  & 1.26 & 135.9 & 0.46 \\ [5pt]
             &             & C1   & 5   & 2.44 & 143.2 & 0.76 \\ [5pt]
             & 1999 Nov 9  & Core & 92  & ...  & ...   & 0.10 \\ [5pt]
             &             & C3   & 13  & 0.47 & 128.6 & 0.24 \\ [5pt]
             &             & C2   & 9   & 1.41 & 137.2 & 0.52 \\ [5pt]
             &             & C1   & 4   & 2.60 & 144.6 & 0.94 \\ [5pt]
             & 2000 Jan 7  & Core & 75  & ...  & ...   & 0.11 \\ [5pt]
             &             & C3   & 14  & 0.45 & 127.6 & 0.23 \\ [5pt]
             &             & C2   & 8   & 1.31 & 137.2 & 0.51 \\ [5pt]
             &             & C1   & 5   & 2.45 & 144.3 & 0.85 \\ [5pt]
             & 2000 Mar 23 & Core & 87  & ...  & ...   & 0.09 \\ [5pt]
             &             & C3   & 16  & 0.44 & 127.9 & 0.23 \\ [5pt]
             &             & C2   & 10  & 1.39 & 137.7 & 0.62 \\ [5pt]
             &             & C1   & 4   & 2.69 & 144.5 & 0.93 \\ \tableline \\
\end{tabular}}
\end{center}
\vspace{-0.15in}
$a$: Flux density in millijanskys. \\
$b$: $r$ and PA are the polar coordinates of the
center of the component relative to the presumed core.
Position angle is measured from north through east. \\
$c$: $a$ is the Full Width at Half Maximum (FWHM) of the circular Gaussian
component.
\end{table*}

We estimate that these full-track observations allow us to measure the positions
of the component centers to within 10\% of a uniform beam width, and this is the error
assumed for subsequent analysis (calculated by taking 10\% of the projection of the 
elliptical beam FWHM onto a line joining the center of the core to the center of the component).
Error bars larger than this produce a scatter about the fits to linear component motion
that are so small as to be statistically unlikely, confirming that our estimated error is reasonable.
Note that the low rms noise achieved by these long integrations allows us to detect even the faintest component
(the 5 mJy C1 in 1ES~2344+514) with an SNR exceeding 50:1.

Because the jets of these three sources appear relatively smooth,
we have also investigated whether the jets can be represented by a smoothly
varying function such as a power law, rather than a series of discrete Gaussians.
We used the method  of Xu et al. (2000), who fit the jets of
FR I radio galaxies with power laws by first subtracting a Gaussian core component,
and then producing a one-dimensional curve of summed flux density across the jet
vs. distance along the jet.  Xu et al. (2000) adequately fit such curves
(which we hereafter refer to as jet `profiles') that they constructed for their sample of radio galaxies
with power laws with an index of about $-$2.

In Figure 4 we first show the jet profiles, including the core component,
obtained from the CLEAN images (the solid curves),
along with the corresponding profiles obtained from the Gaussian components that were fit
to the visibilities (the dotted curves).
The vertical lines indicate the centers of the model-fit Gaussians.
In Figure 5 we show the jet profiles with the core component subtracted
(the solid curves) along with the power law that is the best fit to this jet profile (the dotted curves).
We show these curves for the three sources considered in this paper, and for a 15 GHz observation
of the TeV blazar Mrk~501 from data presented by Edwards \& Piner (2002).
Both smooth power laws (with fitted indices that fall between $-$1 and $-$2) 
and a series of discrete Gaussians that sum to produce
an approximately smooth curve are adequate fits to the observed jet profiles
for PKS~2155$-$304 and 1ES~2344+514.  The jet profile of 1ES~1959+650 does
show a smooth decline, but this curve is not well fit by a power law.
The jet profile of Mrk~501 has local maxima, and would not be well fit by any
monotonic function.  Note that the jet profile of 1ES~2344+514 also shows small local maxima
near the locations of the model-fit Gaussians.

\begin{figure*}[!t]
\plotfiddle{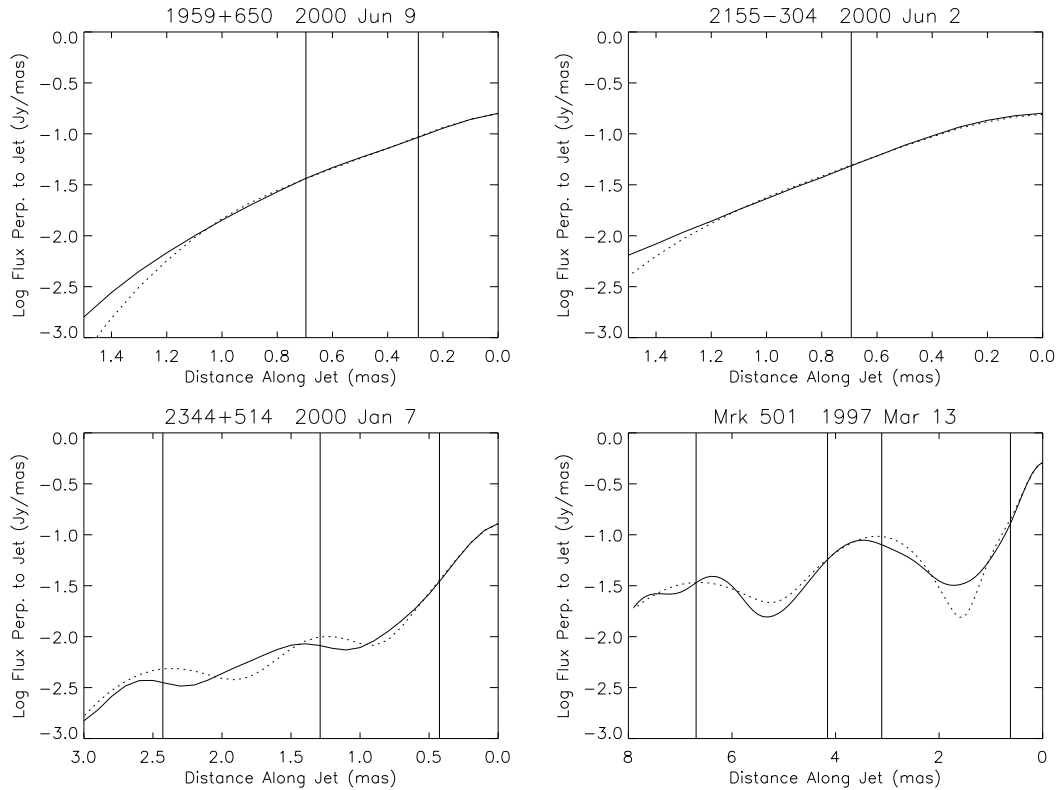}{4.0in}{90}{60}{60}{220}{-30}
\caption{Jet profiles showing the summed flux across the jet as a function of distance
along the jet for 1959+650 on 2000 Jun 9, 2155$-$304 on 2000 Jun 2, 2344+514 on 2000 Jan 7,
and Mrk~501 on 1997 Mar 13.  Data on Mrk~501 are from a 15 GHz observation from Edwards and Piner (2002). The solid
curves show the summed flux from the CLEAN image, the dotted curves show the profiles produced
by the Gaussian models that were fit to the visibility data, from Table~\ref{mfittab}.
The vertical lines show the positions of the centers of the Gaussians that comprise
the Gaussian model.}
\end{figure*}

\begin{figure*}[!t]
\plotfiddle{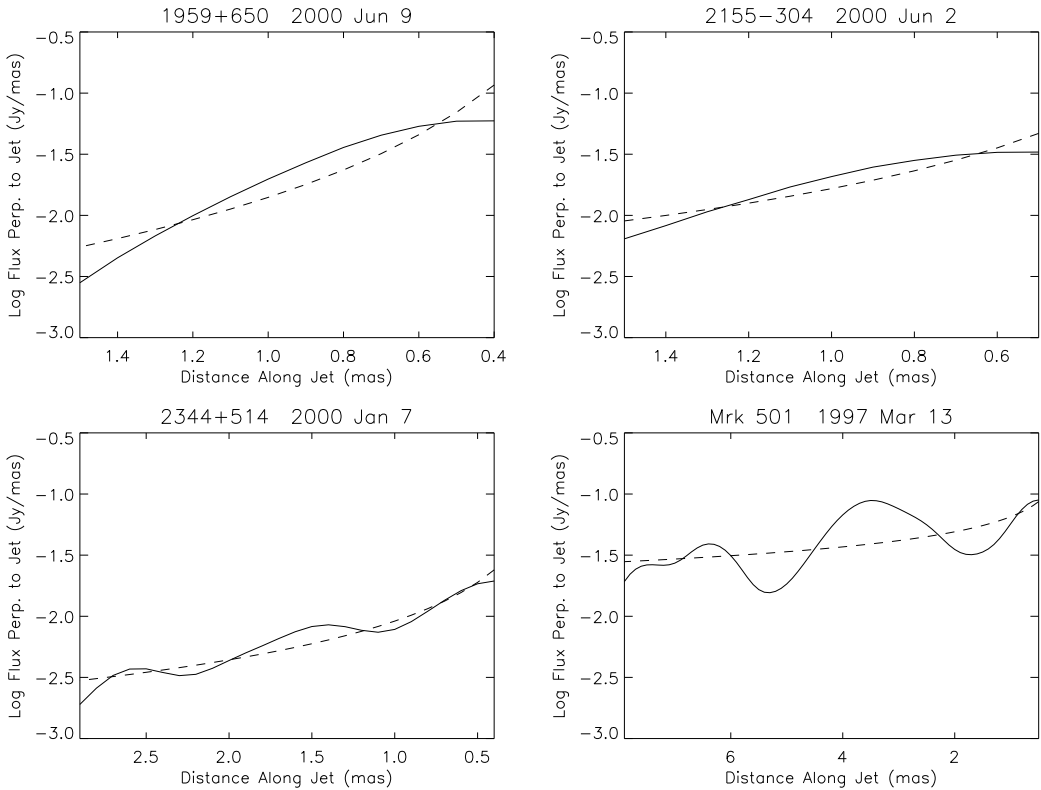}{4.0in}{90}{60}{60}{220}{-30}
\figcaption{Jet profiles showing the summed flux across the jet as a function of distance
along the jet for 1959+650 on 2000 Jun 9, 2155$-$304 on 2000 Jun 2, 2344+514 on 2000 Jan 7,
and Mrk~501 on 1997 Mar 13.  Data on Mrk~501 are from a 15 GHz observation from Edwards and Piner (2002). The core
component has been subtracted from the image prior to the summing.  The solid curves show the
summed flux from the CLEAN image after the subtraction of the core, the dotted curves show
the power law that is the best fit to the solid curve.}
\end{figure*}

There are then two possibilities: It is possible that these three jets are intrinsically a series of discrete components
that are blended into an approximately smooth profile by the 
limited resolution close to the core, it is also possible that the
jet has an intrinsically smooth profile that is being reproduced in the model fitting
by the appropriate sum of discrete Gaussians.  Given the limited
range over which we can follow these low flux density jets, we
can not distinguish between these two possibilities using only
the data from these three sources, and both discrete Gaussians and 
power laws provide about equally good fits to the jet profiles, for two out of the three sources
(1ES~1959+650 is better fit by Gaussians).

To resolve this issue, we return to our observations of the
TeV sources Mrk~421 (Piner et al. 1999) 
and Mrk~501 (Edwards \& Piner 2002), which are very similar to
the three sources studied here except that their
radio flux density is nearly an order of magnitude higher, and their
jets can therefore be followed further from the core and with higher dynamic range.
In Mrk~421 and Mrk~501, it is evident from jet profile plots that
these jets show
a series of local maxima and minima, that are much better fit by Gaussian
components than by a smooth power law.  The profiles of Mrk~501 are
shown in Figures 4 and 5, and plots for Mrk~421 show 
similar structure.  By analogy with the two
better-observed TeV blazars, we retain the Gaussian model for source
morphology in the remainder of this paper, making the assumption that the jets of the fainter 
(in the radio) TeV sources are
similar to the jets of the brighter sources Mrk~421 and Mrk~501.
Representation by Gaussians is also necessary to facilitate comparison of these sources with published VLBI
data on other sources, such as the EGRET blazars (Jorstad et al. 2001a) or VLBA 2~cm survey sources (Kellermann et al. 2000),
as well as with Mrk~421 and Mrk~501.

\section {Results}
\label{results}
\subsection{VLBI Core}
We estimate that the flux densities of the VLBI cores given in Table~\ref{mfittab}
are accurate to within about 10\%, taking into account the corrections made
by amplitude self-calibration, and estimated errors in the model fits.
With 10\% error bars there is no significant detection of core flux density
variability, except in the case of PKS~2155$-$304, for which there is a marginal detection of
variability, with a 0.025 $\chi^{2}$ probability
of constant flux density.  The VLBI core flux density of this source decreases by 30\% between
2000 March and 2000 June, and then increases again by 20\% by the end of 2000 June.

From the core flux densities and measured sizes we can calculate the 
brightness temperature, 
which can be used as an indicator of the amount of Doppler boosting, characterized by the Doppler factor $\delta$.
The maximum brightness temperature of a circular Gaussian is given by
$T_{B}=1.22\times10^{12}\;\frac{S(1+z)}{a^{2}\nu^{2}}$~K,
where $S$ is the flux density of the Gaussian in janskys,
$a$ is the FWHM of the Gaussian in mas,
$\nu$ is the observation frequency in GHz, and $z$ is the redshift.
Observed brightness temperatures are amplified by a factor of $\delta$ over intrinsic brightness temperatures,
and often lie above physical limits such as the inverse Compton limit ($\sim10^{12}$ K, Kellermann \& Pauliny-Toth 1969)
and equipartition limit ($\sim10^{11}$ K, Readhead 1994).
A lower limit on $\delta$ can then be invoked that reduces the observed brightness temperature
below the appropriate limit (e.g., Tingay et al. 2001).

The Gaussian core components in Table~\ref{mfittab} all have similar brightness temperatures of a 
few times 10$^{10}$ K, with a mean brightness temperature of $3\times10^{10}$ K.
The core components are all partially resolved, and lower limits on the core sizes
(and therefore upper limits on the brightness temperature) were obtained using the
Difwrap program for model component error analysis (Lovell 2000).
The brightness temperature upper limits are about twice the best-fit values,
with a mean brightness temperature upper limit of $6\times10^{10}$ K.
The equipartition brightness temperature for these three sources is $\approx6\times10^{10}$ K,
using equation (4a) of Readhead (1994), and therefore no relativistic beaming needs to be invoked
to bring the brightness temperature below either the inverse Compton or equipartition limit.
In fact, if $\delta$ is high in the VLBI core, then the core has a large departure from
equipartition and minimum energy (Readhead 1994).
There is thus no indication from the VLBI core properties that the core emission
is highly beamed, for any of these three sources.

\subsection{VLBI Jet}
Apparent speeds for the VLBI jet components in Table~\ref{mfittab} were measured
by making linear least-square fits to the separation of the components from the VLBI core
versus time.  These fits are shown in Figures~6$a-6c$.
Fitted speeds are given in Table~\ref{speedtab} for the three sources contained in this
paper, as well as for Mrk~421 (Piner \& Edwards in preparation) and Mrk~501 (Edwards \& Piner 2002).
The apparent component speeds are mostly subluminal, and many are consistent with no
motion (stationary components).  The notable exception is PKS~2155$-$304, but the error
bar on this speed measurement is large.
In fact, the measured component speeds for the three
sources studied in this paper, when considered together, are statistically consistent 
with no motion ($\chi^{2}$ probability of 0.12 for no motion of any component).
Because of the slow or non-existent component motions, there are no correlations found
between VLBI component ejections and episodes of high-energy activity, as was found
for the EGRET blazars by Jorstad et al. (2001b).
Again, the one exception may be PKS~2155$-$304, where the 1$\sigma$ range
on the zero-separation epoch of the jet component includes the 1997 November TeV
and X-ray high state, but this error range is large enough that this is not conclusive.

\begin{figure*}
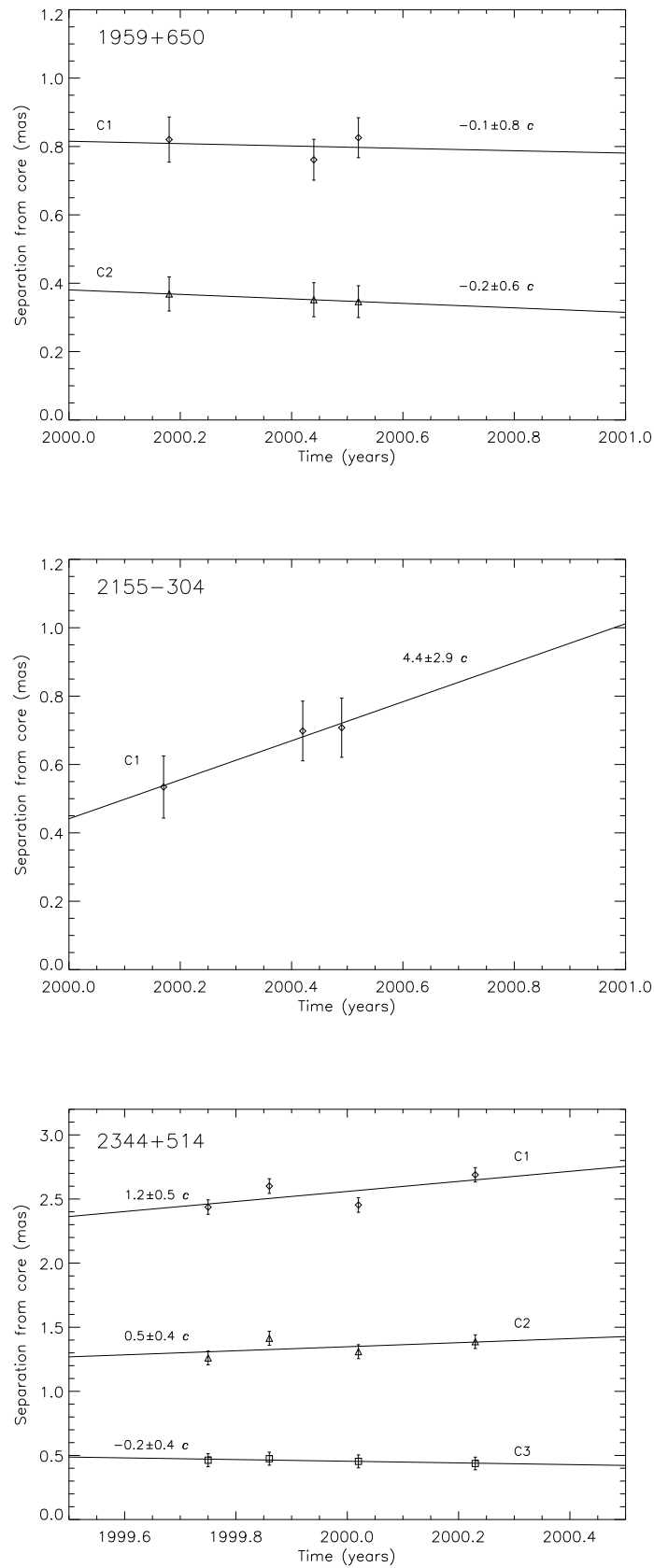

\plotfiddle{f6a.epsi}{8.0in}{90}{40}{40}{150}{390}
\plotfiddle{f6b.epsi}{0.0in}{90}{40}{40}{150}{190}
\plotfiddle{f6c.epsi}{0.0in}{90}{40}{40}{150}{-10}
\caption{($a$)-($c$): Distances from the core of Gaussian component centers as
a function of time.  Error bars are 10\% of the projection of the uniformly-weighted beam along
a line joining the center of the core to the center of the component.
The lines are the least-squares fits to outward motion with constant speed.
The fitted apparent speeds are shown next to these lines.
($a$): 1ES~1959+650  ($b$): PKS~2155$-$304  ($c$): 1ES~2344+514.}
\end{figure*}

\begin{table*}[!t]
\begin{center}
{\small
\caption{Apparent Component Speeds in TeV Blazars}
\label{speedtab}
\begin{tabular}{l c c c c} \tableline \tableline \\[-5pt]
& & Apparent Speed$^{a}$ & & $\theta^{b}$ \\ [5pt]
Source & Comp. & (multiples of $c$) & Ref. & (deg) \\ \tableline \\[-5pt]
Mrk~421      & C4    & $0.04\pm0.06$  & 1 &     \\ [5pt]
             & C5    & $0.20\pm0.05$  & 1 & 0.2 \\ [5pt]
             & C6    & $0.18\pm0.05$  & 1 &     \\ [5pt]
             & C7    & $0.12\pm0.06$  & 1 &     \\ [5pt]
             & C8    & $0.06\pm0.03$  & 1 &     \\ [5pt]
Mrk~501      & C1    & $0.05\pm0.18$  & 2 &     \\ [5pt]
             & C2    & $0.54\pm0.14$  & 2 & 0.6 \\ [5pt]
             & C3    & $0.26\pm0.11$  & 2 &     \\ [5pt]
             & C4    & $-0.02\pm0.06$ & 2 &     \\ [5pt]
1ES~1959+650 & C1    & $-0.11\pm0.79$ & 3 & 0.8 \\ [5pt]
             & C2    & $-0.21\pm0.61$ & 3 &     \\ [5pt]
PKS~2155$-$304 & C1    & $4.37\pm2.88$  & 3 & 4.2 \\ [5pt]
1ES~2344+514 & C1    & $1.15\pm0.46$  & 3 & 1.3 \\ [5pt]
             & C2    & $0.46\pm0.43$  & 3 &     \\ [5pt]
             & C3    & $-0.19\pm0.40$ & 3 &     \\ \tableline
\end{tabular}}
\\ [5pt]
\end{center}
$a$: for $H_{0}=71$ km s$^{-1}$ Mpc$^{-1}$, $\Omega_{m}=0.27$, and $\Omega_{\Lambda}=0.73$.\\
$b$: Angle to the line-of-sight calculated for
an assumed Doppler factor of 10, using the highest measured
component speed (or speed upper limit) for each source,
not meant to be used as the actual angle
to the line-of-sight (see text).
\\
References. --- (1) Piner \& Edwards, in preparation;
(2) Edwards \& Piner (2002)
with modified cosmological parameters;
(3) this paper
\end{table*}

It should be noted that because of the relatively short time period 
spanned by these observations (4 to 6 months, depending on the source),
any relatively slow speed will be statistically consistent with no motion.
There is thus a lower limit to the speed that can be distinguished from a stationary
component with this data, corresponding to $\approx1c$ for 1ES~1959+650 and 1ES~2344+514,
and $\approx4c$ for PKS~2155$-$304 (this higher limit is due to the higher redshift, shorter
time coverage, and more elliptical beam for this source).

\section{Discussion}
\label{discussion}
The parsec-scale jets of the TeV blazars are different in character from the jets
of the other well-studied $\gamma$-ray selected sample, the EGRET blazars, whose
VLBI properties were studied by Jorstad et al. (2001a, 2001b).
A Kolmogorov-Smirnov (KS) test comparing the apparent speeds measured in the
TeV blazars (Table~\ref{speedtab}) to those measured for EGRET blazars (Jorstad et al. 2001a)
shows a difference in the apparent speed distribution with 99.98\% confidence, with
the TeV blazars having slower apparent speeds.  KS tests also show the TeV blazar
apparent speeds to be slower than those in radio-selected samples: comparison with the Caltech-Jodrell Bank Flat Spectrum (CJF) Survey
(Vermeulen 1995) shows a difference with 99.91\% confidence, and comparison with the 2 cm survey (Kellermann et al. 2000)
shows a difference with a somewhat lower 97.90\% confidence.

The presence of stationary components in the TeV blazars is not surprising, because in all of the
VLBI surveys mentioned above, one-third to one-half of the VLBI components observed are found to be very slow
or stationary.  However, superluminally moving components are usually also present in the jets of these sources
with stationary components, indicating that the jet is not intrinsically slow (Jorstad et al. 2001a).
What causes the statistical difference in the speed distribution of the TeV blazars is not the presence of 
stationary components, but the lack of any superluminally moving components.
This lack of any superluminally moving features is the same conclusion that would be reached
if we had used the smooth power law fits to the jets ($\S$~\ref{modelfits}), but representation by Gaussians
is required for the quantitative comparison given above.  The important question to be answered about the
parsec-scale jets of these sources is then:  What has become of the relativistically moving shocks that are assumed
to be responsible for the high-energy flaring activity, and that are clearly visible in the jets of the EGRET blazars?

The first hypothesis that should be examined is that the moving components are present, but appear slow
due to a very small angle to the line-of-sight.  The angle to the line-of-sight calculated from the 
highest measured speed (or speed upper limit) for each source is given in Table~\ref{speedtab}, for an
assumed $\delta$ of 10 (a typical lower limit from models of the multiwavelength spectra and variability).
These angles to the line-of-sight are typically less than a degree.
Even though $\gamma$-ray selected samples should have a distribution that lies closer to the line-of-sight
than radio-selected samples (because of the dependence of the $\gamma$-ray flux on a higher power of $\delta$, 
see the Monte Carlo simulations of Lister 1998),
they should not typically be so close to the line-of-sight as to appear subluminal, and the Monte Carlo
simulations of Lister (1998) predict a faster apparent speed distribution for $\gamma$-ray selected samples, 
for both SSC and ERC $\gamma$-ray emission.
While the extreme misalignments on parsec scales, or between parsec and kiloparsec scales, in three of these
sources (Mrk~501, 1ES~1959+650, and PKS~2155$-$304) 
does indicate a fairly small angle to the line-of-sight, the typical angle is more likely to be
a few degrees, rather than less than one degree.
Note also that Giebels et al. (2002) state that the ``quiescent'' X-ray emission from these sources can be explained
by jets aligned within a few degrees of the line of sight.

Based on other indications that $\delta$ in the parsec-scale radio jet is actually low (such as the
low brightness temperatures), we consider the more likely explanation for the sluggish parsec-scale jets to be that 
the bulk Lorentz factor has been reduced between the TeV emitting sub-parsec scale and the parsec scale.
Such a change in the bulk Lorentz factor (from $\Gamma>10$ to $\Gamma$ of a few) also provides an explanation
for problems encountered in BL Lac -- FR I unification (Georganopoulos \& Kazanas 2003).
We note that the jet decollimation observed in all of these sources at a few mas from the core 
is also indicative of a jet that has little momentum left, and is easily influenced by the
external medium.  This morphology in Mrk~501 is interpreted by Giroletti et al. (2003) in terms
of a decelerating two-component (fast inner spine and slower outer layer) jet model, based on flux profiles transverse to the jet.  
For the weaker sources presented in this paper, we do not have sufficient sensitivity or resolution transverse to the jet
to attempt decomposition into separate spine and layer components.
The lack of any observed counterjets in our images, with a lower limit on the jet to counterjet brightness ratio
$J > \sim 100$, limits the amount of deceleration that has occurred, and constrains the bulk Lorentz factor to 
remain $> \sim 2$ on parsec scales.
If the jets of TeV blazars are decelerating,
the important question then becomes: What is the mechanism for jet deceleration, and why
does it work more efficiently in the TeV blazars than the EGRET blazars?

Most scenarios for the production of TeV $\gamma$-ray flares (and superluminal VLBI components) are based
on `shock-in-jet' models.  In these models, shocks transfer bulk kinetic energy to internal energy of the plasma,
which then radiates.  The shock models can be broadly divided into two categories: internal shocks
(e.g., Spada et al. 2001, Sikora \& Madejski 2000) where the shocks are due to interactions of different portions of the jet with
varying densities or bulk speeds, and external shocks (e.g., Dermer \& Chiang 1998),
where the jet plasma interacts with density variations in the external medium.
The efficiency of the model is a measure of how effectively the shocks transfer bulk kinetic 
energy from the jet.  Current internal shock models for blazars (Spada et al. 2001, Tanihata et al. 2003)
have low efficiency (e.g., Figure~3 of Spada et al. 2001 shows deceleration from $\Gamma\approx25$ to $\Gamma\approx20$
on parsec scales).  However, low efficiency was built into the Spada et al. (2001) model in order to reproduce
bulk relativistic motion on parsec scales, by enforcing small differences in bulk Lorentz factors between
colliding shells.  Tanihata et al. (2003) suggest that the efficiency can be increased by making this
difference larger, but that there may then be problems with reproducing the observed flare timescales.
The efficiency of deceleration in the Dermer \& Chiang (1998) model can be quite high, e.g., their Figure~2
shows deceleration of a plasmoid to mildly relativistic speeds on sub-parsec scales.
Regardless of which shock mechanism is dominant,
we suggest that the deceleration of the jet to relatively small bulk Lorentz factors at
parsec scales has now been established as an observed property of TeV blazars,
and that this property of TeV blazar jets should now be used to constrain the shock models for these sources.

\section{Conclusions}
This paper is part of a series exploring the parsec-scale jet structure
of TeV blazars through multi-epoch VLBI observations.  In Piner et al. (1999) we presented
observations of Mrk~421, in Edwards \& Piner (2002) we presented observations of Mrk~501,
and in this paper we presented VLBA observations of the fainter (in the radio) TeV blazars
1ES~1959+650, PKS~2155$-$304, and 1ES~2344+514.  All of these sources are quite similar
in their VLBI properties, and our major conclusions from this study to date are:
\begin{enumerate}
\item{All of these sources have similar VLBI morphologies.  There is an initial collimated
jet extending to a few mas from the core, that may appear strongly bent.  Beyond this region
the jet loses collimation and transitions to a diffuse, low surface brightness morphology.} 
\item{The superluminally moving shocks or `components' present in other blazars
are not apparent in the parsec-scale jets of these sources.  The components in TeV blazar
jets are predominantly stationary or subluminal.}
\item{The VLBI cores of the three sources studied here are partially resolved and have brightness temperatures
of a few $\times10^{10}$ K (compared with a few $\times10^{11}$ K in the case of Mrk~421 and Mrk~501).  
A high Doppler factor is not required to reduce the observed
brightness temperatures below the equipartition value.}
\item{Counterjets are not observed, and the lower limit placed on the jet to counterjet
brightness ratio is $J > \sim 100$.}
\end{enumerate}

Based on these four points, we conclude that the jets of TeV blazars are only mildly
relativistic ($\Gamma\sim2-4$) on parsec-scales.  We suggest the following scenario for the evolution of TeV blazar jets.
The jets start out highly relativistic, as required by models reproducing
the multiwavelength spectra and variability.
Internal or external shocks transfer bulk kinetic energy from the jet with a high efficiency, such that
the jet becomes only mildly relativistic a few parsecs from the core.
The bulk Lorentz factor has decreased substantially, and the shocks have dissipated, by the time the jet reaches 
the parsec scales that we are observing with VLBI, including the VLBI core out to a few mas from the core.
Immediately beyond this inner jet region, the jet undergoes rapid decollimation, 
probably due to a low-momentum jet interacting with the external environment.
This decollimation is evident on the VLBA images of all of these sources beyond a few mas from the core.
The challenge is now to see if shock-in-jet models can produce deceleration on the scales 
observed here while also reproducing the other observed multiwavelength properties 
of these sources, and to explain why the TeV blazars apparently decelerate their jets more efficiently
than do the EGRET blazars.

We are currently analyzing additional VLBA data on the TeV blazars, including
dual-circular-polarization observations of Mrk~421 after its 2001 TeV high state,
and observations to further clarify the jet structures in 1ES~1959+650 and PKS~2155$-$304,
as well as in the recently detected TeV blazar H~1426+428.

\acknowledgments
Part of the work described in this paper has been carried out at the Jet
Propulsion Laboratory, California Institute of Technology, under
contract with the National Aeronautics and Space Administration.
The National Radio Astronomy Observatory is a facility of the National Science Foundation operated
under cooperative agreement by Associated Universities, Inc.
This research has made use of
the NASA/IPAC Extragalactic Database (NED) which is operated by the Jet Propulsion Laboratory, California Institute of
Technology, under
contract with the National Aeronautics and Space Administration. 
B.~G.~P. acknowledges support from the NASA
Summer Faculty Fellowship Program.


\begin{references}

Aharonian, F.~A. et al. 2002, A\&A, 353, 847

Aharonian, F., et al. 2003a, A\&A, 403, L1

Aharonian, F., et al. 2003b, A\&A, 406, L9

Beckmann, V., Wolter, A., Celotti, A., Costamante, L., Ghisellini, G.,
Maccacaro, T., \& Tagliaferri, G. 2002, A\&A, 383, 410

Bennett, C.~L., et al. 2003, ApJ, in press (astro-ph/0302207)

Bondi, M., Marcha, M.~J.~M., Dallacasa, D., \& Stanghellini, C. 2001, MNRAS, 325, 1109

Catanese, M., et al. 1998, ApJ, 501, 616

Chadwick, P.~M., et al. 1999, ApJ, 513, 161

Chiappetti, L., et al. 1999, ApJ, 521, 552

Dermer, C.~D., \& Chiang, J. 1998, New Astronomy, 3, 157

Djannati-Ata\"{\i}, A., et al. 2003, in Active Galactic Nuclei: from Central Engine to Host Galaxy,
ed. S. Collin, F. Combes, \& I. Shlosman (San Francisco:ASP), 291

Djannati-Ata\"{\i}, A., et al. 2003, preprint (astro-ph/0307452)

Edelson, R., Griffiths, G., Markowitz, A., Sembay, S., Turner, M.~J.~L., \&
Warwick, R. 2001, ApJ, 554, 274

Edwards, P.~G., \& Piner, B.~G. 2002, ApJ, 579, L67

Gaidos, J.~A., et al. 1996, Nature, 383, 319

Georganopoulos, M., \& Kazanas, D. 2003, ApJ Letters, in press (astro-ph/0307404)

Ghisellini, G. 2003, preprint (astro-ph/0306429)

Giebels, B., et al. 2002, ApJ, 571, 763

Giommi, P., Padovani, P., \& Perlman, E. 2000, MNRAS, 317, 743

Giroletti, M., et al. 2003, ApJ, submitted

Holder, J., et al. 2003a, ApJ, 583, L9

Holder, J., et al. 2003b, preprint (astro-ph/0305577)

Horan, D., et al. 2002, ApJ, 571, 753

Horan, D., et al. 2003, preprint (astro-ph/0305578)

Itoh, C., et al. 2002, A\&A, 396, L1

Jorstad, S.~G., Marscher, A.~P., Mattox, J.~R., Wehrle, A.~E.,
Bloom, S.~D., \& Yurchenko, A.~V. 2001a, ApJS, 134, 181

Jorstad, S.~G., Marscher, A.~P., Mattox, J.~R., Aller, M.~F., Aller, H.~D.,
Wehrle, A.~E., \& Bloom, S.~D. 2001b, ApJ, 556, 738

Kataoka, J., Takahashi, T., Makino, F., Inoue, S., Madejski, G.~M., Tashiro, M.,
Urry, C.~M., \& Kubo, H. 2000, ApJ, 528, 243

Kellermann, K.~I. \& Pauliny-Toth, I.~I.~K. 1969, ApJ, 155, L71

Kellermann, K.~I., Vermeulen, R.~C., Zensus, J.~A., \& Cohen, M.~C. 2000,
in Astrophysical Phenomena Revealed by Space VLBI, ed. H. Hirabayashi, P.~G. Edwards, \&
D.~W. Murphy (Sagamihara:ISAS), 159

Laurent-Muehleisen, S.~A., Kollgaard, R.~I., Moellenbrock, G.~A., \& Feigelson, E.~D.
1993, AJ, 106, 875

Lister, M.~L. 1998, Ph.D. thesis, Boston Univ.

Lovell, J.~E.~J. 2000, in Astrophysical Phenomena Revealed by Space VLBI, ed. H. Hirabayashi, P.~G. Edwards, \&
D.~W. Murphy (Sagamihara:ISAS), 301

Marscher, A.~P. 1999, Astroparticle Physics, 11, 19

Nishiyama, T., et al. 2000, in AIP Conf. Proc. 516, Proc. 26th International Cosmic Ray Conference,
ed. B.~L. Dingus, D.~B. Kieda, \& M.~H. Salamon (Melville:AIP), 370

Perlmutter, S., et al. 1997, ApJ, 483, 565

Piner, B.~G., Unwin, S.~C., Wehrle, A.~E., Edwards, P.~G., Fey, A.~L., \& Kingham, K.~A. 
1999, ApJ, 525, 176

Piner, B.~G., Edwards, P.~G., Fodor, S., \& Rector, T.~A. 2002, Publ. Astron. Soc. Australia, 19, 114 

Readhead, A.~C.~S. 1994, ApJ, 426, 51

Rector, T.~A., Gabuzda, D.~C., \& Stocke, J.~T. 2003, AJ, 125, 1060

Shepherd, M.~C., Pearson, T.~J., \& Taylor, G.~B. 1994, BAAS, 26, 987

Sikora, M., \& Madejski, G. 2000, ApJ, 534, 109

Spada, M., Ghisellini, G., Lazzati, D., \& Celotti, A. 2001, MNRAS, 325, 1559

Tanihata, C., Urry, C.~M., Takahashi, T., Kataoka, J., Wagner, S.~J., Madejski, G.~M., Tashiro, M., \& Kouda, M.
2001, ApJ, 563, 569

Tanihata, C., Takahashi, T., Kataoka, J., \& Madejski, G.~M. 2003, ApJ, 584, 153

Tingay, S.~J., et al. 2001, ApJ, 549, L55

Tommasi, L., Diaz, R., Palazzi, E., Pian, E., Poretti, E., Scaltriti, F., \& Treves, A.
2001, ApJS, 132, 73

Vermeulen, R.~C. 1995, Proc. Natl. Acad. Sci., 92, 11385

Villata, M., Raiteri, C.~M., Popescu, M.~D., Sobrito, G., De Francesco, G., Lanteri, L., \&
Ostorero, L. 2000, A\&AS, 144, 481

Xie, G.~Z., Zhou, S.~B., Dai, B.~Z., Liang, E.~W., Li, K.~H., Bai, J.~M., Xing, S.~Y., \&
Li, W.~W. 2002, MNRAS, 329, 689

Xu, C., Baum, S.~A., O'Dea, C., Wrobel, J.~M., \& Condon, J.~J. 2000, AJ, 120, 2950

Zhang, Y.~H., et al. 2002, ApJ, 572, 762

\end{references}
\end{document}